\documentclass{article}
\usepackage{spconf,amsmath,graphicx,booktabs,caption,multirow,bbm}
\usepackage{makecell} 

\title{AN AUDIO-TEXTUAL DIFFUSION MODEL FOR CONVERTING SPEECH SIGNALS INTO ULTRASOUND TONGUE IMAGING DATA}
\name{Yudong Yang$^{1}$, Rongfeng Su$^{1,*}$\thanks{*Corresponding authors: Rongfeng Su(email: rf.su@siat.ac.cn); Lan Wang(email: lan.wang@siat.ac.cn).}, Xiaokang Liu$^{1,2}$, Nan Yan$^{1,3}$, Lan Wang$^{1,3,*}$}
\address{$^{1}$Shenzhen Institute of Advanced Technology, Chinese Academy of Sciences, Shenzhen, China.\\
$^{2}$University of Chinese Academy of Sciences, Beijing, China.\\
$^{3}$Guangdong-Hong Kong-Macao Joint Laboratory of Human-Machine Intelligence-Synergy Systems, \\
Shenzhen, China
}

\begin{document}
%
\maketitle
\begin{abstract}
Acoustic-to-articulatory inversion (AAI) is to convert audio into articulator movements, such as ultrasound tongue imaging (UTI) data. An issue of existing AAI methods is only using the personalized acoustic information to derive the general patterns of tongue motions, and thus the quality of generated UTI data is limited. To address this issue, this paper proposes an audio-textual diffusion model for the UTI data generation task. In this model, the inherent acoustic characteristics of individuals related to the tongue motion details are encoded by using wav2vec 2.0, while the ASR transcriptions related to the universality of tongue motions are encoded by using BERT. UTI data are then generated by using a diffusion module. Experimental results showed that the proposed diffusion model could generate high-quality UTI data with clear tongue contour that is crucial for the linguistic analysis and clinical assessment. The codes and examples can be found on the website\footnote{https://yangyudong2020.github.io/wav2uti/}.
\end{abstract}
\section{Introduction}
\label{sec:intro}
\begin{sloppypar}
The problem of estimating the movement of articulators from speech signals is known as acoustic-to-articulatory inversion (AAI). In recent years, AAI has garnered increasing attention due to its potential applications in speech technology, such as automatic speech recognition~\cite{8114358,9640504}, speech therapy~\cite{ARENS20212869,RIBEIRO202124}, and speech assessment~\cite{9383619,10096920}. The articulatory movements can be measured through various techniques, for instance real-time magnetic resonance imaging (rt-MRI)~\cite{10094797}, X-ray microbeam~\cite{8114358}, electromagnetic articulography (EMA)~\cite{9640504}, and ultrasound tongue imaging (UTI)~\cite{8851769,7471970}. Compared to the more invasive EMA and the operationally complex rt-MRI, UTI has the characteristics of simple operation, non-invasive, and clear visualization. 
The tongue contour in UTI data is the most important part to display the tongue motor function~\cite{utim}.
Suitable AAI methods to generate UTI data with clear tongue contour is of great importance in clinical applications.

The current AAI methods can be divided into two categories: discriminative and generative methods. Discriminative methods generate the sequence data of the tongue by identifying the mapping relationship between the speech signal of each pronunciation and the corresponding tongue motions~\cite{8114358,8851769,10094703}. Due to the lack of modelling ability regarding the temporal dependencies across multiple consecutive pronunciations, the quality of generated tongue motions is limited. For example, the state-of-the-art DNN based AAI system generated UTI data with blurry tongue contour~\cite{8851769}
, which is difficult to obtain tongue motion trajectories for further applications. In contrast, generative methods learn the joint probability distribution of speech and articulatory motions, which can generate results with better temporal coherence~\cite{10094797,kim23g_interspeech,scheck23_interspeech}. Researchers hoped to directly derive the general patterns of tongue movements from large amounts of personalized acoustic inputs by using various neural network structure~\cite{8851769, 7471970}. For the UTI data generation task in this paper, it is difficult to acquire high-quality UTI data from limited speech-ultrasound parallel training data in many real applications. In fact, the tongue motions related to the tongue contour in UTI data can be divided into the individual and universal parts. The individual part means the personal details in tongue movements, such as the height of the tongue position. The universal part means the general patterns of tongue movements, which should be consistent for the same consecutive pronunciations of different speakers. Obviously, additional text inputs contain rich universal information with respect to the tongue movements, which may reduce the requirement of parallel training data volume and thus improve the quality of generated UTI data.

In this study, inspired by the successful use of diffusion model in medical image generation tasks with richer details~\cite{reynaud2023feature,kim23g_interspeech,yang2022diffusion,N1}, a multimodal diffusion model that integrates speech and textual information is proposed for the UTI data generation task. To best of our knowledge, this study is the first to apply the diffusion model for converting audio into UTI data. The proposed model consists of two stages: conditional encoding and UTI data generation. In the first stage, the inherent acoustic characteristics of individuals related to the details of tongue movements are encoded by using wav2vec2.0~\cite{baevski2020wav2vec}, while the ASR transcriptions in the textual space related to the universality of tongue motions are encoded by using BERT~\cite{devlin2018bert}. The personalized acoustic information and universal textual information are then fused by using cross-attention mechanism. In the second stage, the fused features are used as the conditional inputs of the diffusion module~\cite{220403458}. In this module, the long-term dependencies between consecutive pronunciations and UTI data are captured by using cyclic denoising sampling strategy. Experimental results on a Mandarin speech-ultrasound dataset showed that the proposed audio-textual diffusion AAI system outperformed the state-of-the-art DNN-based AAI system by a LPIPS~\cite{zhang2018unreasonable} improvement of 67.95\% relative, with a FID~\cite{heusel2017gans} decreased from 256.80 to 22.02. This demonstrates that the generated results exhibit data distribution and diversity closer to real UTI data. Besides, the proposed diffusion AAI system generated high-quality UTI data with clear tongue contour, which is crucial for clinical assessment tasks, such as tongue function assessment.

\end{sloppypar}

\section{METHODS}
\label{sec:format}
\begin{sloppypar}
In this section, audio-only diffusion model and the audio-textual diffusion model will be presented for generating ultrasound tongue imaging (UTI) data from speech signals.
\vspace{-13pt}  
\subsection{Audio-only Diffusion Model for UTI Generation}
\label{ssec:subhead}
\vspace{-3pt}  
The audio-only diffusion model for UTI data generation is illustrated in Fig.1(a). The original speech inputs were first pre-processed by the Wav2vec2.0~\cite{baevski2020wav2vec} to obtain the acoustic embeddings for describing different pronunciations. The acquired acoustic embeddings $c$ and the corresponding time steps $t$ were then fed into the UTI data generation module.

The core of the UTI data generation module is the diffusion model, which is shown in Fig.I. The generic pipeline of diffusion models involves a forward process and a reverse process to learn a data distribution, as well as a sampling procedure to generate novel data. For the forward process, supposed that $x$ is the UTI data, $q_{utl}(x)$ represents the distribution of real UTI data and the corresponding standard deviation is $\sigma_{utl}$. By adding a i.i.d Gaussian noise with the standard deviation $\sigma$ to the UTI data, a family of distributions $p(x;\sigma)$ can be obtained. When $\sigma_{max}\gg\sigma_{utl}$, $p(x;\sigma_{max})$ is approximated to the Gaussian noise. The key point of the diffusion model is to randomly sample a noisy start point $x_0\sim \mathcal{N}(0,\sigma_{max}^2\mathbf{I})$,  and then denoise it into the UTI data $x_i$ with noise level $\sigma_{max}=\sigma_0>\sigma_1>\cdots>\sigma_N=0$, where $x_i\sim p(x;\sigma_i)$.  The finally generated UTI data $x_N$ will have the approximated distribution of real data $q_{utl}(x)$.

In order to reverse the forward process of the diffusion model and gradually restore to the true data distribution from a purely random state. A denoising function $D(x,\sigma)$ is introduced to minimize the $L2$ denoising error of all extracted samples from $q_{utl}$, the loss function is defined as follow:
\begin{equation}
\setlength{\abovedisplayskip}{3pt}
\setlength{\belowdisplayskip}{3pt}
\mathcal{L}=\mathbbm{E}_{y\sim q_{utl}}\mathbbm{E}_{n\sim \mathcal{N}\left(0,\sigma_{max}^2\mathbf{I}\right)}||D\left(y+n,\sigma\right)-y{||}_2^2
\end{equation}
Where $y$ is a training data point and $n$ is noise. By using the ordinary differential equations (ODE)~\cite{karras2022elucidating}, the noise level of each data point can be increased or decreased by moving it forward or backward in the diffusion process, respectively. The requirement of ODE is satisfied by:
\begin{equation}
\setlength{\abovedisplayskip}{3pt}
\setlength{\belowdisplayskip}{3pt}
    dx=-\sigma\left(t\right)\sigma\left(t\right)\nabla_x\log{p\left(x;\sigma\left(t\right)\right)}dt
    \label{eqdx}
\end{equation}
where the dot denotes a time derivative, $\sigma\left(t\right)=t$ is a schedule, $\nabla_x\log p(x;\sigma)$ is the score function~\cite{Song2021Score-Based}, a vector field that points towards higher density of data at a given noise level. In this paper, the tongue contour in UTI data is supposed to be a higher density vector field. According to the equation~(\ref{eqdx}) used in the denoising process, the network optimization will pay more attention to the field related to the tongue contour in UTI data, and thus the UTI data with clearer tongue contour can be acquired.

The aim of this study is to generate coherent and natural UTI data under given speech conditions. First, the low-quality UTI data sequence $v_0$ is generated from a pretrained speech encoder, such as Wav2vec2.0 used in this paper. Then, $v_0$ is used as a conditional input for the next diffusion model, which follows the concept of a Cascade Diffusion Model to generate a high-quality and clearer $v_1$. As shown in Fig.1(c), compared to traditional image diffusion models, the architecture based on Unet~\cite{220403458} incorporates changes in different levels and adds time-aware layers through 3D convolutions. We use the raw speech signal $V_{audio}$ as a condition with the Elucidated Diffusion Model (EDM)~\cite{karras2022elucidating} setup, and apply it to UTI data generation. The denoising models in the cascade are defined as $D_{\theta_s}$, where $s$ defines the rank of the cascaded models, and $D_{\theta_0}$ is defined as:
\setlength{\belowdisplayskip}{3pt}
\begin{align}
D_{\theta_0}\left(x;\sigma,\lambda_c\right) &= c_{skip}\left(\sigma\right)x \notag \\
&+ c_{out}\left(\sigma\right)F_{\theta_s}\left(c_{in}\left(\sigma\right)x;c_{noise}\left(\sigma\right),\lambda_c\right)
\end{align}
Where $F_{\theta_s}$ is the neural network transformation of $D_{\theta_0}$ and $D_{\theta_0}$ outputs $v_0$. Supposed that $\sigma_{utl}$ is the standard deviation of the real UTI data distribution, $c_{skip}$, $c_{out}$, $c_{in}$ and $c_{noise}$ are defined as:
\setlength{\belowdisplayskip}{3pt}
\begin{align}
c_{skip}\left(\sigma\right)=\left(\sigma_{q_{utl}}^2\right)/({\sigma^2+\sigma}_{q_{utl}}^2) \\
c_{out}\left(\sigma\right)=\sigma\ast\sigma_{q_{utl}}/{({\sigma^2+\sigma}_{q_{utl}}^2)}^{0.5}                \\
c_{in}\left(\sigma\right)=1/{({\sigma^2+\sigma}_{q_{utl}}^2)}^{0.5}       \\ 
c_{noise}\left(\sigma\right)=\log(\sigma_t)/4    
\end{align}
\begin{figure*}[htbp]
\centering
\includegraphics[width=7.1in, keepaspectratio]{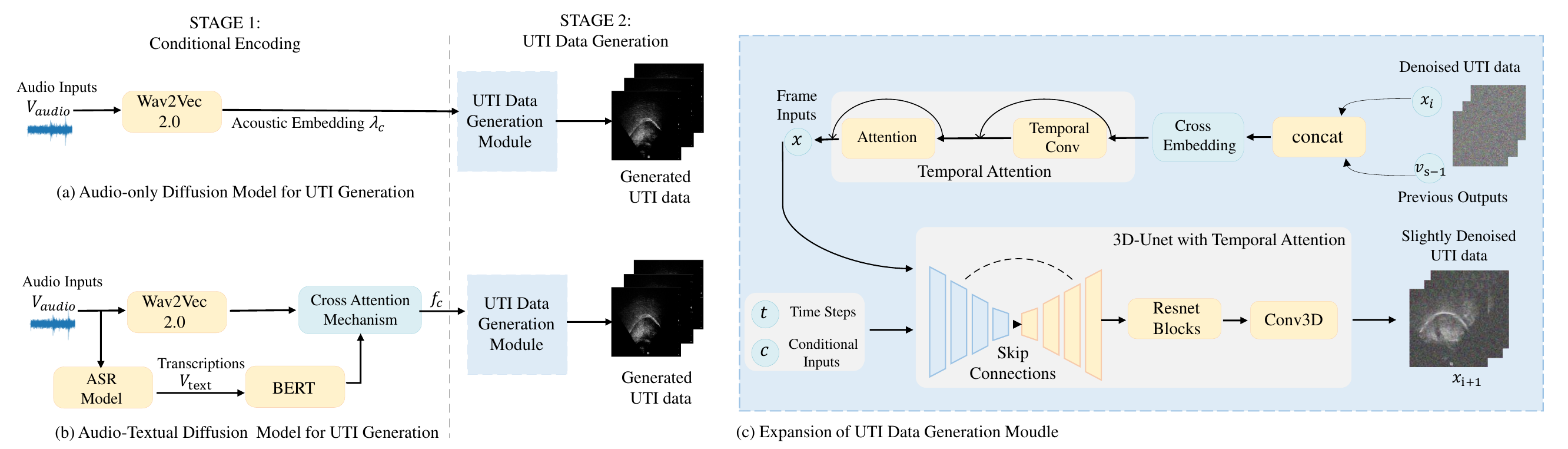}  
\caption{The overall structure of the audio-only and audio-textual diffusion models for ultrasound tongue imaging (UTI) data generation, where $f_c$ is the fusion conditions from the acoustic and textual encoding.}
\vspace{-15pt}  
\end{figure*}
\vspace{-12pt}

The sampling procedure is accomplished through a random sampling method. The sampling begins with a noise sample $x_0\sim \mathcal{N}(0,t_0^2\mathbf{I})$ , and the noise level is given by $\sigma\left(t_i\right)=t_i$, Then, we calculate the variable factor $\gamma_i$, which is used to control the magnitude of noise level adjustment $\sigma\left(t_i\right)$ over the time step $t_i$. The variable factor $\gamma_i$ is defined as:
\begin{equation}
\setlength{\abovedisplayskip}{3pt}
\setlength{\belowdisplayskip}{3pt}
\gamma_i(t_i)=min(s_{churn}/N,\sqrt{2}-1)\forall t_{i}\in(s_{tmin},s_{tmax})
\end{equation}
Where $N$ represents the number of sampling steps.  For $\forall i\in\left\{0,\ldots,N-1\right\}$, we sample the $\epsilon_i\sim \mathcal{N}(0,S_{noise}\mathbf{I})$ and calculate the slightly increased noise level ${\hat{t}}_i=\left(\gamma_i\left(t_i\right)\right)+1)t_i$. The previous sample is ${\hat{x}}_i=x_i+{({\hat{t}}_i^2-t_i^2)}^{0.5}\epsilon_i$. Then, this sample is applied in the denoising model $D_\theta$ and the local slope is $d_i=({\hat{x}}_i-D_\theta({\hat{x}}_i;{\hat{t}}_i))/{\hat{t}}_i$, which is used to predict the next sample $x_{i+1}={\hat{x}}_i+({\hat{t}}_{i+1}-{\hat{t}}_i)d_i$. Except for the last step, a correction to $x_{i+1}$ is applied: $d_i^\prime=(x_{i+1}-D_\theta(x_{i+1};t_{i+1}))/t_{i+1}$ and $x_{i+1}={\hat{x}}_i+(t_{i+1}-{\hat{t}}_i){(d}_i+d_i^\prime)/2$. This process is then iterated sequentially in the cascade model, using the previous cascade output $v_{s-1}$ as a condition, and having the speech encoding $\lambda_c$.
\vspace{-10pt}  
\subsection{Audio-Textual Diffusion Model for UTI  Generation}
\label{ssec:subhead}
\vspace{-3pt}  
As discussed before, for the given consecutive pronunciations, the tongue motions can be divided into individual detailed information and universal movement patterns. In fact, the personalized speech signals contain rich individual detailed information related to the tongue motions, such as the tongue position. The text inputs, as the speaker-independent semantic representation, inherently contain the generality information about tongue movements. In this paper, we hope to improve the model ability to generate clear and coherent UTI data by integrating the personalized detailed acoustic information and the universal textual information.

The proposed audio-textual diffusion model is shown in Fig. 1(b). In the proposed model, the audio inputs were fed into the large Wav2vec 2.0 to obtain the acoustic embeddings. To obtain reliable text inputs from the speech signals, a well-trained large Whisper ASR model is used in this paper. This ASR model achieved a character error rate (CER) of 0.02\% on the whole Mandarin speech-ultrasound dataset. Inspired by the success of BERT~\cite{devlin2018bert} applied in language understanding tasks, the ASR transcriptions was preprocessed by a well-trained BERT to derive the word embeddings. Such word embeddings are supposed to contain the high-level semantic information that can reflex the universality of the tongue motions.
The acquired acoustic and textual embeddings were then fused by using the cross-attention mechanism~\cite{rombach2022high}. The fused features $f_c$ were used as the conditional inputs $c$ of the UTI data generation module, as shown in Fig. 1(c). The training process of the audio-textual diffusion model is similar to the one described in Sec. 2.1. The only difference is $D_{\theta_0}$ of equation (2). Supposed that the audio inputs are $V_{audio}$ and the ASR transcriptions are $V_{text}$,  $D_{\theta_0}$ is defined as:
\begin{align}
\begin{split}
D_{\theta_0}\left(x;\sigma,f_c\right) &= c_{skip}\left(\sigma\right)x+ \\
&c_{out}\left(\sigma\right)F_{\theta_s}\left(c_{in}\left(\sigma\right)x;c_{noise}\left(\sigma\right),f_c\right)
\end{split}
\end{align}
\end{sloppypar}
\vspace{-0pt}  
\section{EXPERIMENT}
\label{sec:pagestyle}
\subsection{Datasets}
\label{ssec:subhead}
\begin{sloppypar}
\vspace{-5pt}  
Experiments were conducted on Mandarin speech-ultrasound dataset. This dataset was collected from 44 healthy persons with three different speech tasks (vowel, word and sentence), totally 6.85 hours. The training set consists of 40 speakers, while the test set consists of 4 speakers. No overlap exists between the training and test sets. The UTI data were recorded in mid sagittal orientation using a Focus\&Fusion Finus 55 ultrasound system with a sampling rate of 60 fps and a resolution of 920$\times$700. The P5-2 phased array probe was fixed by an ultrasound stabilization headset. The speech data were recorded by a BOYA BY-WM4 PRO microphone with a sampling frequency of 16khz and single channel. The speech signals and UTI data were synchronized by using an external sound card. The downsampled UTI data under different resolutions were used for training different models.
\vspace{-10pt}  
\subsection{Implement Details}
\label{ssec:subhead}
\vspace{-3pt}  
Following the state-of-the-art AAI system based on DNN for the UTI data generation task~\cite{8851769}, a similar DNN-based AAI baseline system was reimplemented in this paper. All neural networks were trained on four NVIDIA A6000 GPUs with a batch size of 4. The experimental setup included hyperparameters with a minimum noise level of 0.002, a maximum noise level of 160, a data distribution standard deviation of 0.25, and a learning rate of $5\times 10^{-4}$. All neural networks were trained with $5\times 10^{6}$ iterations, and the parameters were sampled every 1000 iterations. For the audio part, the large Wav2Vec2.0 were obtained from a 56k-hour corpus. For the text part, the well-trained large BERT has 110M parameters. During the training process of all neural networks, the parameters of both large Wav2Vec2.0 and large BERT  were frozen.
\vspace{-25pt}  
\subsection{Evaluation Metrics}
\label{ssec:subhead}
\vspace{-5pt}  
Various   metrics are used to comprehensively assess the quality of generated UTI data. The root Mean Square Error (RMSE) and Peak Signal-to-Noise Ratio (PSNR) metrics are used to evaluate the  pixel-level similarity and quality differences. Besides, Learned Perceptual Image Patch Similarity(LPIPS) and Fréchet Inception Distance (FID) are used to evaluate the perception quality of UTI data and the similarity of statistical distributions, which are computed using Seqeeze~\cite{simonyan2014very} and I3D~\cite{skorokhodov2022stylegan} networks, respectively.
\vspace{-12pt}  
\subsection{Experimental Result}
\label{ssec:subhead}
\vspace{-3pt}  
The performance of the DNN-based AAI system and the proposed diffusion AAI system is shown in Table 1. It could be observed that the proposed AAI system consistently outperformed the DNN-based AAI baseline in all metrics (RMSE, LPIPS, PSNR and FID). For example, a LPIPS improvement of 67.95\% relative and a FID improvement of 91.43\% relative were obtained from the diffusion system using audio-textual information over the DNN-based AAI baseline. The generated results are displayed in Fig. 2.
\begin{table}[htbp]
\fontsize{9}{11}\selectfont
\label{table2}
\vspace{-9pt}  
\caption{Performance comparison of the DNN-based AAI baseline system and the proposed diffusion AAI system under a UTI output size of 112×112. “A” represents the audio-only inputs, and “A+T” represents the audio-textual inputs.}
\vspace{-5pt}  
\centering
\scalebox{0.96}{
\begin{tabular}{@{}p{1.8cm}cccccc@{}}
\toprule
Models & Inputs & RMSE↓ & PSNR↑ & LPIPS↓ & FID↓ \\
\midrule
DNN~\cite{8851769} & A & $5.8489$ & $32.8234$ & $0.3323$ & $256.80$ \\
\midrule
\multirow{2}{*}{\centering Ours} & A & $5.6635$ & $33.1081$ & $0.1206$ & $27.81$ \\
& A+T & $\textbf{5.6341}$ & $\textbf{33.1482}$ & $\textbf{0.1065}$ & $\textbf{22.02}$ \\
\bottomrule
\end{tabular}
}
\vspace{-6pt}  
\end{table}

We also compared the performance of the proposed diffusion AAI system with or without additional textual information under different UTI output resolutions. As shown in Table 2, the diffusion AAI system using textual inputs consistently outperformed the counterpart without textual inputs. For example, even if under the largest UTI output size of 112*112, the audio-textual AAI system outperformed the audio-only AAI system by a LPIPS reduction of 11.69\% relative and a FID reduction of 20.82\% relative, respectively. The results in Table 2 have given the strong evidences that the introduction of additional textual information can significantly enhance the quality of the generated UTI data, although the used speech-ultrasound  training data is only 6.39 hours. 
\begin{table}[htbp]
\centering
\caption{Performance comparison of the diffusion AAI systems with or without additional textual information under various UTI output resolutions.}
\vspace{-5pt}  
\label{table3}
\fontsize{9}{12}\selectfont
\centering
\scalebox{1}{
\begin{tabular}{@{}m{1.55cm}cccccc@{}}
\toprule
\#Resolution & Inputs & RMSE↓ & PSNR↑ & LPIPS↓ & FID↓ \\
\midrule
\multirow{2}{*}{\centering $64\times64$} & A & $5.8114$ & $32.8734$ & $0.0972$ & $80.51$ \\
& A+T  & $5.7512$ & $32.9820$ & $0.0946$ & $78.65$ \\
\midrule
\multirow{2}{*}{\centering $96\times96$} & A & $5.7681$ & $32.9355$ & $0.1547$ & $50.30$ \\
& A+T & $5.7025$ & $33.0411$ & $0.1195$ & $37.01$ \\
\midrule
\multirow{2}{*}{\centering $112\times112$} & A & $5.6635$ & $33.1081$ & $0.1206$ & $27.81$ \\
&  A+T & $\textbf{5.6341}$ & $\textbf{33.1482}$ & $\textbf{0.1065}$ & $\textbf{22.02}$ \\
\bottomrule
\end{tabular}
}
\vspace{-10pt}  
\end{table}

In order to intuitively illustrate the performance of different AAI systems, some examples of the UTI data generated from the DNN-based AAI baseline and the proposed diffusion systems are shown in Fig. 2. As shown in Fig. 2, compared to the other AAI systems, the diffusion AAI system using additional textual information can generated the best UTI data with the most clear tongue contour.
\vspace{-9pt}  
\begin{figure}[h]
\centering
\includegraphics[width=0.96\linewidth, keepaspectratio]{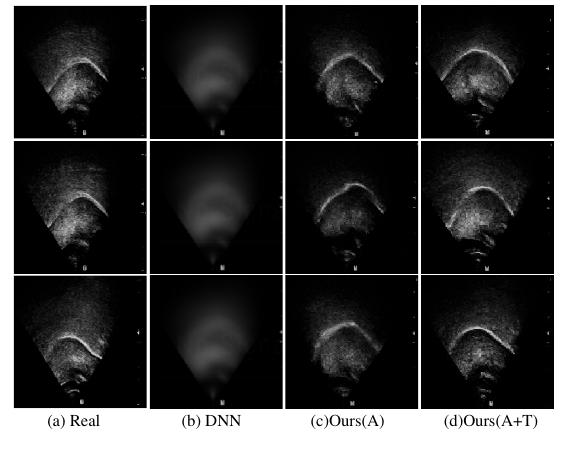}  
\vspace{-15pt}  
\caption{Examples of real and generated UTI data.}
\vspace{-15pt}  
\end{figure}
\end{sloppypar}
\vspace{-3pt} 
\section{CONCLUSION}
\label{sec:majhead}
\vspace{-3pt} 
\begin{sloppypar}
In this paper, the diffusion model is first applied for converting audio into ultrasound tongue imaging (UTI) data. Using additional textual inputs, the proposed diffusion AAI system can generate high quality UTI data with clear tongue contour, which is crucial for clinical assessment tasks, such as tongue function assessment. Future works will focus on the new applications of diffusion model related to speech orders. 
\end{sloppypar}
\vspace{-3pt} 
\section{ACKNOWLEDGEMENTS}
\label{sec:majhead}
\vspace{-3pt} 

This work is supported by National Natural Science Foundation of China (U23B2018, NSFC 62271477), Shenzhen Science and Technology Program (JCYJ20220818101411025, JCYJ20220818101217037, JCYJ20220818102800001), and Shenzhen Peacock Team Project (KQTD20200820113106007).

\bibliographystyle{IEEEbib}
\bibliography{refs}

\end{document}